\documentclass[useAMS,usenatbib,usegraphicx]{mn2e}

\usepackage{amssymb}

\title[Non-thermal neutrinos from supernovae leaving a magnetar]{Nonthermal neutrinos from supernovae leaving a magnetar}
\author[S. Horiuchi et al.]
{Shunsaku Horiuchi,$^{1}$\thanks{E-mail:
horiuchi@utap.phys.s.u-tokyo.ac.jp}
Yudai Suwa,$^{1}$ Hajime Takami,$^{1}$ Shin'ichiro Ando$^{2}$
\newauthor
and Katsuhiko Sato$^{1,3}$\\%
$^{1}$Department of Physics, School of Science, The University of Tokyo, 
Tokyo 113-0033, Japan\\%
$^{2}$California Institute of Technology, Mail Code 130-33, Pasadena, CA 
91125, USA\\%
$^{3}$Institute for the Physics and Mathematics of the Universe, The University
of Tokyo, Kashiwa, Chiba, 277-8582, Japan}

\begin{document}

\date{\today}

\pagerange{\pageref{firstpage}--\pageref{lastpage}} \pubyear{2008}

\maketitle

\label{firstpage}

%latex maths
%\newcommand{\gtrsim}{\lower.5ex\hbox{$\; \buildrel > \over \sim \;$}}
%\newcommand{\lesssim}{\lower.5ex\hbox{$\; \buildrel < \over \sim \;$}}

%Journals
\newcommand\apj{ApJ}
\newcommand\apjl{ApJ}
\newcommand\apjs{ApJS}
\newcommand\nat{Nature}
\newcommand\mnras{MNRAS}
\newcommand\aap{A\&A}
\newcommand\physrep{Physics Reports}

\begin{abstract}%
Under the fossil field hypothesis of the origin of magnetar magnetic fields, the magnetar inherits its magnetic field from its progenitor. We show that during the supernova of such a progenitor, protons may be accelerated to $\sim$10$^4$ GeV as the supernova shock propagates in the magnetic stellar envelope. Inelastic nuclear collisions of these protons produce a flash of high-energy neutrinos arriving a few hours after thermal (10 MeV) neutrinos. The neutrino flash is characterized by energies up to $O(100)$ GeV and durations seconds to hours, depending on the progenitor: those from smaller Type Ibc progenitors are typically shorter in duration and reach higher energies compared to those from larger Type II progenitors. A Galactic Type Ib supernova leaving behind a magnetar remnant will yield up to $\sim$160 neutrino induced muon events in Super-Kamiokande, and up to $\sim$7000 in a km$^3$ class detector such as IceCube, providing a means of probing supernova models and the presence of strong magnetic fields in the stellar envelope.
\end{abstract}
 

%==============================================================================
\section{Introduction}

Since being first predicted by \citet{BaadeZwicky1934}, neutron stars have been observed to display a wealth of phenomena. Over the past decades a new class of neutron stars has emerged, through the studies of the emission mechanism of soft $\gamma$-ray repeaters (SGRs) and anomalous X-ray pulsars (AXPs). Termed magnetars, they are powered by their extreme magnetic fields, typically of the order $10^{14}$--$10^{15}$ G, rather than their spin-down energy loss as in the case of pulsars \citep{1992AcA....42..145P,1992Natur.357..472U,1992ApJ...392L...9D,1993ApJ...408..194T,1995MNRAS.275..255T,1996ApJ...473..322T} \citep[for an overview, see, e.g.,][]{2001ApJ...561..980T,2006RPPh...69.2631H}. Now strengthened by observational measurements including their slow spin periods, fast spin-down rates \citep{1998Natur.393..235K,2006csxs.book..547W} and spectral properties \citep{2002Natur.419..142G,2003ApJ...584L..17I,2003ApJ...586L..65R}, there is increasing evidence for their extreme magnetic fields \citep[for a review of observations, see, e.g.,][]{2006csxs.book..547W}. 
 
The origin of the magnetic field however remains debated. Of the magnetar candidates---5 SGRs, 6 AXPs, and a few radio pulsars---approximately a third are associated within known young supernova remnants, suggestive of an origin in massive star explosions \citep{2006csxs.book..547W} \citep[but see also][]{2001ApJ...559..963G}. The ages of the remnants are $\sim$10$^4$ years, consistent with the inferred ages of magnetars derived from their spin-down rates (``spin-down age''), which lie tightly between $10^3$ and $10^4$ years. If magnetars are young neutron stars resulting from core-collapse supernovae, then their magnetic fields could have been inherited from their progenitors. This is the so-called fossil field hypothesis, where magnetic flux is conserved and the field is amplified during the core-collapse process \citep{2006RPPh...69.2631H,2006MNRAS.367.1323F}. Observations reveal that there is, in principle, enough magnetic flux, i.e., the magnetic flux of O-stars, derived from their recently detected magnetic fields \citep{2002MNRAS.333...55D,2006MNRAS.365L...6D,2008MNRAS.387L..23P}, are comparable to those of magnetars. On the other hand, the magnetic field may be generated by a convective dynamo in the first \emph{O}(10) seconds of the protoneutron star birth \citep{1993ApJ...408..194T}. In this process, the energy in differential rotation is converted to magnetic energy. It is not yet clear which is the dominant process.

Here we discuss proton acceleration and production of high-energy neutrinos during a supernova which leaves behind a magnetar. We focus on the fossil field hypothesis, which endows the star with strong stellar magnetic fields. We explore three progenitors corresponding to Type Ic, Ib and II supernovae, and show that proton acceleration is realised in all the cases during the propagation of the shock through the stellar envelope. The maximum proton energy is sufficiently high that neutrinos are produced through inelastic proton-proton collisions. The neutrinos have energies of 0.1 GeV up to $O(100)$ GeV, which is significantly higher than thermal neutrinos from the collapsed core ($\sim$10 MeV). The signal thus resides in a unique energy window, a positive point from a detection perspective.

The neutrino emission and its detectability depends on the energy loss rate of muons and pions. We find that a supernova explosion at the Galactic Centre results in 70--160 neutrino induced muon events in Super-Kamiokande, with the largest value for a Type Ib supernova. Higher energy neutrinos in the sensitivity range of km$^3$ class detectors are more strongly dependent on the progenitor, with 200--6600 events per supernova. The progenitor dependency is the product of two physical processes, proton acceleration and energy loss of pions and muons. For Type Ic supernovae, the compact progenitor results in strong cooling of pions and muons, so that high-energy neutrino emission is suppressed. The cooling is dominated by inverse-Compton scattering on electron synchrotron photons. For Type II supernovae, the maximum neutrino energy is intrinsically low. The greatest emission of high-energy neutrinos is realised for a Type Ib supernova. 

In all the cases, the neutrinos arrive tens of hours after thermal neutrinos, and last between seconds to hours depending on the progenitor radius. The background atmospheric neutrino is significantly smaller, and detection is essentially background free in most cases. Given the nature of the fossil field hypothesis, detection of these high-energy neutrinos can provide support for the stellar origin of magnetar magnetic fields. Detection also provides useful diagnostics for supernova properties.

The paper is structured as follows. In section 2 we discuss the progenitor density profile and the supernova shock environment. These are the required background for investigating proton acceleration in the shock, which we address in section 3. We also include the production of neutrinos and their detection prospects in section 3. In section 4 we finish with discussions. Throughout this paper we define $Q_\alpha = Q / 10^\alpha$ for a quantity $Q$ in cgs units. The exception is $\epsilon_x$, the energy of particle species $x$, for which we use GeV.

%==============================================================================
\section{Setup}

In this paper we investigate proton acceleration at a supernova shock propagating through the stellar envelope. Both proton acceleration and particle cooling depend on quantities near the shock front, such as the magnetic field, photon density, and particle density. We therefore start with a discussion of these required quantities, and treat proton acceleration in the next section. Here, we discuss the progenitor density profile (section 2.1), the progenitor magnetic field (section 2.2), and the particle density and temperature around the supernova shock (section 2.3).

%------------------------------------------------------------------------------
\subsection{Stellar density}

First we discuss our selection of progenitor density profiles. It is widely believed that neutron stars are formed from the collapse of massive OB stars, with main-sequence mass ranging between $8 \lesssim M/ \mathrm M_{\sun} \lesssim 45$. We assume that magnetars are similarly produced as remnants of core-collapse supernova of massive stars. Due to the lack of observational evidence for which type of core-collapse preferentially produces a magnetar remnant,\footnote{However, we note that a connection between magnetar birth and type Ic supernova has been proposed in the context of low-luminosity $\gamma$-ray bursts \citep{2006Natur.442.1018M,2007ApJ...659.1420T}.} we consider three supernova progenitors: a CO star, a He star, and a blue supergiant (BSG), which respectively give rise to a Type Ic, Ib and II supernova. 

\begin{figure}
\includegraphics[width=84mm]{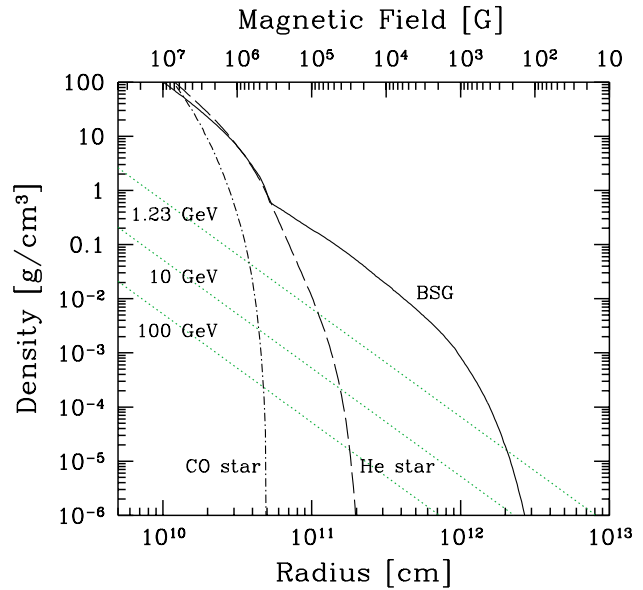}
\caption{\label{parameter}Plot showing ``acceleration possible'' (below dotted lines) and ``no acceleration'' (above dotted lines) regions, where the numbers by the dotted lines indicate the maximum accelerated proton energy. The density profiles of our chosen models are also plotted: CO star (resulting in Type Ic supernova, dot-dashed), He star (resulting in Type Ib supernova, dashed), and BSG (resulting in Type II supernova, solid). The supernova shock traces the density curves. Therefore, acceleration becomes possible as the shock approaches the stellar surface.}
\end{figure}

For the CO star model (Type Ic progenitor), we adopt the presupernova model 16SK from \citet{2006ApJ...637..914W}, motivated by the $\gamma$-ray burst association with core-collapse supernovae. This is a rapidly rotating star with solar metallicity and initial mass of $16\,\mathrm M_{\sun}$. Due to strong mass loss, the final mass is $M_*\approx 5 \, \mathrm M_{\sun}$ and radius is $R_* \approx 5 \times 10^{10}$ cm. The outer envelope is radiative and dominated by carbon. 

For the He star model (Type Ib progenitor), we adopt the presupernova model 12SE from \citet{2006ApJ...637..914W}. This is a rapidly rotating star with solar metallicity and initial mass of $12 \, \mathrm M_{\sun}$. Due to mass loss, the final mass is $M_*\approx 7\, \mathrm M_{\sun}$ and radius is $R_* \approx 2 \times 10^{11}$ cm. The outer envelope is radiative and dominated by helium. 

For the BSG model (Type II progenitor) we adopt the non-rotating presupernova model 16TA from \citet{2006ApJ...637..914W}. This is a low metallicity star with an initial mass of $16 \, \mathrm M_{\sun}$. There is little mass loss, and the final mass is $M_* \approx 16 \, \mathrm M_{\sun}$ and radius is $R_*\approx 3 \times 10^{12}$ cm. The outer envelope is radiative and dominated by hydrogen. Note that we only consider BSGs which have radiative envelopes, and do not consider more common red supergiants. This is because radiative envelopes are necessary for survival of fossil magnetic fields \citep{2003A&A...403..693M,2004MNRAS.355L..13T}.

In Fig.~\ref{parameter} we show the density profiles of our adopted models. Throughout this paper, when we provide numerical values to equations, we adopt the He star model and the radius $0.9 R_*$, for illustrative purposes. The illustrative density is $\approx$10$^{-5} \, \mathrm{g\,cm^{-3}}$.

%------------------------------------------------------------------------------
\subsection{Stellar magnetic field}

Now we discuss the progenitor magnetic field. We work under the fossil origin for the magnetic field of neutron stars. As we discuss below, this requires the progenitor to be strongly magnetized.

The magnetic fields of massive stars have recently been directly detected. The measured strengths of surface dipole fields are as high as 1 kG. Although at the present time there are only a handful of such direct detection, including $\theta^1$ Ori C with $1.1 \pm 0.1$ kG \citep{2002MNRAS.333...55D}, HD191612 with 1.5 kG \citep{2006MNRAS.365L...6D}, and two in the Orion Nebula with $1100^{+320}_{-200}$ G and $650^{+220}_{-170}$ G \citep{2008MNRAS.387L..23P}, it has been speculated that magnetism may be widespread among massive stars \citep{2008MNRAS.387L..23P}. Observed properties of the magnetic field, such as its global nature and no clear correlation with stellar parameters, favour a fossil origin over a dynamo origin; i.e.,~the magnetic fields are fossil remnants from the star formation stage, as relics of the field that pervaded the interstellar medium \citep{2006MNRAS.370..629D}.

Interestingly, the magnetic flux of $\theta^1$ Ori C, calculated from its observed magnetic field and inferred radius \citep{2006A&A...448..351S} to be $(7 \pm 3) \times 10^{27} \,\mathrm{G\,cm^2}$, is remarkably close to that of the highest field magnetar SGR 1806-20, $\sim$$3\times 10^{28}\,\mathrm{G\,cm^2}$ (assuming a radius $10^6$\,cm). Therefore, in principle there is enough magnetic flux present in a massive magnetized star to explain the magnetic fields of magnetars, and hints at a possible evolutionary link. This is the so-called fossil field hypothesis of neutron stars, where magnetic flux conservation results in field amplification during collapse. In addition to the assumption that magnetic flux is conserved, it is implied that the field must somehow survive the post-main-sequence evolution and the various internal structural changes during the formation of the neutron star. Despite these ideal assumptions, the scenario could provide a powerful explanation for the wide range of magnetic fields present in neutron stars \citep{2003A&A...403..693M,2004MNRAS.355L..13T,2005MNRAS.356..615F,2006MNRAS.367.1323F}.

It is worth commenting that the fossil field hypothesis of neutron stars is an extension of the fossil theory for magnetic white dwarfs. In the case of white dwarfs, the fraction of the strongly magnetized population is roughly compatible with the fraction of magnetism in early-type stars from which white dwarfs arise \citep{2005MNRAS.356.1576W}. The broad study of magnetic fields in massive OB stars has recently started \citep{2008MNRAS.387L..23P}, which would enable similar studies for neutron stars.

Under the fossil field hypothesis, the neutron star inherits its magnetic field from the progenitor. First we define a mass cut, $M_\mathrm{cut} \lesssim 2\, \mathrm M_{\sun}$, corresponding to the requirement that the collapsed object forms a neutron star and not a black hole, i.e., $M_\mathrm{cut}$ is the collapsed mass, while the overlaying mass $M_*-M_\mathrm{cut}$ is ejected by the supernova shock. The magnetic field of the remnant object must therefore originate from the material within the mass $M_\mathrm{cut}$. In our adopted progenitor models, the iron core mass is $M_\mathrm{Fe} \approx 1.5 \, \mathrm M_{\sun}$, which approximately corresponds to the required mass cut. We therefore assume $M_\mathrm{cut}=M_\mathrm{Fe}$ for simplicity (the exact location is not critical for our purposes). Thus, the magnetic field at the surface of the iron core is found to be, from conservation of magnetic flux,
\begin{equation} \label{CoreBfield}
B_\mathrm{core}=10^{15} \left( \frac{R_{\mathrm{Fe}}}{10^6\,\mathrm{cm}} \right)^{-2} = 10^{11} \,\mathrm{G},
\end{equation}
where $R_{\mathrm{Fe}} \approx$ a few $\times 10^8$ cm is the radius of the iron core.

The magnetic field in the stellar interior, including the envelope, is largely unknown. Following this difficulty, we parameterize the field strength according to a power-law with index $n$,
\begin{equation} \label{Bfield}
B(r)=B_\mathrm{core} \left( \frac{r}{R_{\mathrm{Fe}}} \right)^{-n} \,\mathrm{G}
\quad \mathrm{for}\,\,R_{\mathrm{Fe}} < r < R_*,
\end{equation}
where we have taken the normalization at the iron core radius. For a dipole field, $n=3$. In the fossil theory for neutron stars that we adopt throughout the preset paper, $n=2$ is obtained by equating Eq.~(\ref{Bfield}) to the magnetic field strength measured at the stellar surface. We note that for our illustrative model, the He star, the resulting magnetic field is $\approx$$10^{4.5}$ G at radius $0.9 R_*$.

%------------------------------------------------------------------------------
\subsection{Shock environment}

As the shock propagates through the star, it accumulates stellar material, shocking it. Across the shock, material is compressed; the downstream (labelled $d$) density is $\rho_d=\kappa \rho_u$, where $\rho_u$ is the stellar (upstream) density discussed in section 2.1, and $\kappa$ is the compression ratio. We adopt the fiducial value of $\kappa = 4$, but note that in principle $\kappa$ can reach up to 7 for a radiation shock.

For compact stars, the supernova shock can reach velocities of $\sim$$0.1 c$ as it leaves the stellar surface \citep{1999ApJ...510..379M}. Similar velocities are also seen in numerical studies of explosions of O-Ne-Mg cores, where the steep density gradient drives a fast shock velocity as high as $10^{10}\,\rm{cm\,s^{-1}}$ \citep{2006A&A...450..345K}. Since steep density gradients are generically expected near surfaces of stars, in the current paper we adopt $v_s = 10^{9.5}\,\rm{cm\,s^{-1}}$.

In the optically thick interior of the progenitor, the supernova shock is dominated by radiation \citep{1976ApJS...32..233W,1999ApJ...510..379M}. The radiation
temperature is approximately given by the relation $aT_r^4 \approx \rho_u v_s^2/2$, where $a=4\sigma/c$ and $\sigma$ is the Stefan-Boltzmann constant, so that
\begin{equation} \label{temperature}
T_r \approx 800 \rho^{1/4}_{u,-5}v^{1/2}_{s,9.5} \,\mathrm{eV}.
\end{equation}
The energy densities of the photon and magnetic fields are given as $U_r \approx \rho_u v_s^2/2$ and $U_B=B^2/8\pi$, respectively. Quite generally, $U_r \gg U_B$; this implies the relative importance of inverse-Compton over synchrotron process.

%==============================================================================
\section{Neutrinos from Supernova Shocks}

%------------------------------------------------------------------------------
\subsection{Proton acceleration}

In this section we discuss proton acceleration by diffusive shock acceleration, which produces a power-law spectrum $dN/dp \propto p^{-s}$ for relativistic protons, where $p$ is the particle momentum \citep{1983RPPh...64.1429T,1987PhR...154....1B,2001RPPh...64.1429T}. The acceleration time is $t_\mathrm{acc} \approx 10 D/v_s^2$, where $D=\eta r_g/3$ is the diffusion coefficient of particles close to the shock. Here, $r_g$ is the Larmor radius, and $\eta$ is a factor which depends on the ratio of energy in the ordered magnetic field to that in the turbulent magnetic field. Although there is some uncertainty regarding this coefficient, due to the high level of turbulence close to the shock front, we assume diffusion in the Bohm limit ($\eta \to 1$). We note that this is the most efficient limit for acceleration. We then obtain (in the relativistic regime)
\begin{equation} \label{acceleration}
t_\mathrm{acc}\approx \frac{10}{3} \frac{r_g}{v_s^2} 
= 1 \times 10^{-5}\, \epsilon_{p,1}v^{-2}_{s,9.5}B^{-1}_{4.5}\,\mathrm{s},
\end{equation}
for our fiducial parameters, where $\epsilon_p$ is the proton energy in GeV (we reserve $E_p$ for the total energy in accelerated protons). In the plasma, the proton acquires energy from the bulk plasma motion through resonances \citep{1981Longair}. The plasma waves taking part in this interaction are caused by oscillating electrons. If the photon-electron collision rate exceeds the plasma frequency, these waves will not be produced. Fortunately, we can show that in the radiation dominated shock near the stellar surface, the photon-electron collision rate is smaller than the plasma frequency: while the photon-electron collision rate is $\nu_{\gamma e}=cn_\gamma\sigma_T = 3 \times 10^8 \,\mathrm{s^{-1}}$, the plasma frequency is $\nu_{pl}=2\times 10^{13} \,\mathrm{s^{-1}}$. Here, $\sigma_T$ is the Thomson scattering cross section and $n_\gamma \approx 0.33 (U_r/c \hbar)^{3/4}$ is the photon density,
\begin{equation} \label{PhotonDensity}
n_\gamma \approx 1 \times 10^{22} \rho_{u,-5}^{3/4}v_{s,9.5}^{3/2}
\,\mathrm{cm^{-3}}.
\end{equation}

Proton acceleration must compete against the shortest of three time scales, namely $i$) proton escape from the accelerating region, $ii$) proton energy loss time, and $iii$) age of the accelerator. In supernova shocks, acceleration is limited by the proton energy loss time. 

Accelerated protons lose energy by interacting with the dense photon field and dense protons in the shock vicinity. We first discuss proton cooling by inelastic collisions with target protons. We adopt the target proton density $\kappa \rho_u/m_p$ ($m_p$ is the proton mass), the proton-proton collisional cross section $\sigma_{pp}\approx 4\times10^{-26}\,\mathrm{cm^{2}}$ \citep{Eidelman:2004wy}, and that $20 \%$ of the proton energy is lost in each collision \citep{Bellandi:1994dy}. The cooling time, defined as $\epsilon_p/ |\mathrm{d}\epsilon_p/\mathrm{d}t|$, is then
\begin{equation}
t_{pp} = \frac{1}{0.2 c \, \sigma_{pp} \kappa \rho_u /m_p} 
\approx 2 \times 10^{-4} \rho_{u,-5}^{-1}\,\mathrm{s}.
\end{equation} 

Protons may also lose energy by pair-production on the photon field, also known as the Bethe-Heitler (BH) process. The pairs are produced at rest in the centre of mass frame, and the energy lost by the proton is $\Delta \epsilon_p=2m_e c^2 \gamma_{cm}$, where $\gamma_{cm}=(\epsilon_p + \epsilon_\gamma)/(m_p^2 c^4 + 2 \epsilon_p \epsilon_\gamma)^{1/2}$. The energy loss rate is then $\mathrm{d}\epsilon_p/\mathrm{d}t=n_\gamma c \sigma_{BH} \Delta \epsilon_p$, where $\sigma_{BH}=\alpha r_e^2 [ (28/9)\mathrm{ln}(2\epsilon_p \epsilon_\gamma/m_p m_e c^4)-106/9]$ is the cross section. Here, $\alpha$ and $r_e$ are the fine structure constant and classical electron radius. BH cooling occurs only for protons above $\epsilon_p \approx 5 \times 10^3 \rho_{u,-5}^{-1/4}v_{s,9.5}^{-1/2}$ GeV. Its time scale, determined assuming the average photon energy $\bar{\epsilon}_\gamma \approx 2.7 T_r$ and substituting Eq.~(\ref{PhotonDensity}), reaches a minimum near $\epsilon_p \approx 10^5$ GeV, but is slower than the collisional cooling time.

Finally, protons may cool by synchrotron and inverse-Compton processes with the magnetic and photon fields, respectively. However they are slow, with cooling times of 
\begin{eqnarray}
&&t_{\rm sync}=\frac{3m_p^4 c^3}{4 \sigma_T m_e^2 \epsilon_p U_B}
\approx 4\times 10^8 \epsilon_{p,1}^{-1}B_{4.5}^{-2}\,\mathrm{s}, \\
&&t_{\rm IC}=\frac{3m_p^4 c^3}{4 \sigma_T m_e^2 \epsilon_p U_r}
\approx 400 \epsilon_{p,1}^{-1}\rho_{u,-5}^{-1}v_{s,9.5}^{-2}\,\mathrm{s},
\end{eqnarray}
respectively ($m_e$ is the electron mass).

We conclude from the above analysis that proton-proton collisional cooling dominates the cooling of protons. Equating this cooling time to the acceleration time yields the maximum proton energy,
\begin{equation} \label{maxenergy}
\epsilon_{p,\rm{max}}=\left[ \left( \frac{v_s^2 q
B}{0.2c\sigma_{pp}\kappa\rho_u/m_p} \right)^2 + m_p^2 c^4
\right]^{1/2},
\end{equation}
which is $\epsilon_{p,\rm{max}} \approx 160$ GeV for our fiducial He star and radius $0.9 R_*$. Note that once the shock leaves the star, the proton density is expected to fall dramatically, and cooling will most likely be dominated by pair-production and radiation processes. Inside the star however, Eq.~(\ref{maxenergy}) may be safely used. From equation Eq.~(\ref{maxenergy}) one can identify ``acceleration possible'' regions and ``no acceleration'' regions on the $B$--$\rho_u$ plane. We show this in Fig.~\ref{parameter}, making use of Eq.~(\ref{Bfield}) to plot the radius on the horizontal axis. We plot three threshold lines corresponding to $\epsilon_{p,\rm{max}}=$ 1.23, 10, and 100 GeV. We also show the stellar density profiles adopted. We see that proton acceleration becomes possible as the shock nears the stellar surface. 

Finally, we add a note regarding the possible effects of shock-accelerated electron synchrotron photons. Given the large magnetic field, the energy in accelerated electrons will be rapidly converted to synchrotron photons. An increase in the photon density can potentially cause two problems. First, the photon-electron collision rate may exceed the plasma frequency. Second, proton cooling by inverse-Compton will become faster. We find that for typical energy fractions of relativistic electrons (less than a few percent of the shock energy $E_\mathrm{exp}$ \citep{2001ApJ...558..739A,2003ApJ...589..827B,Waxman:2001kt}), the increase in photon density is not sufficient to cause these problems. For example, substituting $U_r = \xi_e E_\mathrm{exp}/V$ where $V$ is volume, and conservatively assuming $\xi_e=0.01$, the inverse-Compton cooling time scale time is
\begin{equation}
t_{\mathrm{IC},e} \approx 60 \epsilon_{p,1}^{-1} R_{*,11.3}^3 \xi_{e,-2}^{-1} E_{\rm exp,51}^{-1} \,\mathrm{s},
\end{equation} 
which is much slower than collisional cooling. Moreover, these conclusions remains unchanged even if the energy fraction in electrons is maximised (i.e.,~the unrealistic case where $\xi_e = 1$). We therefore safely ignore this effect for protons. Note that it is however important for pions and muons, which we discuss below.

%------------------------------------------------------------------------------
\subsection{Neutrino production}

Inelastic proton-proton interactions producing pions occur for protons with energies above the threshold energy $\epsilon_{p,\rm{th}}=[\frac{1}{2}(m_p+m_n+m_\pi)^2 - m_p^2]/m_p = 1.23$ GeV. The decay of charged pions produce neutrinos through $\pi^+ \to \mu^+ \nu_\mu$ and $\pi^- \to \mu^- \bar{\nu}_\mu$. The muon neutrino energy is $\epsilon_\nu \approx 0.25 \epsilon_\pi \approx 0.05 \epsilon_p$, since $\epsilon_\pi \approx 0.2 \epsilon_p$. Competing with this decay is pion cooling by radiative and collisional processes. We define them similarly as we did for protons, but by adopting $\sigma_{\pi p}= 5\times 10^{-26}\,\mathrm{cm^2}$ \citep{Eidelman:2004wy} and that $80\%$ of the pion energy is lost in each pion-proton collision \citep{Brenner:1981kf}. We find that radiative cooling dominates, in particular inverse-Compton scattering on electron synchrotron photons. Equating the dominant cooling time to the decay time, $\tau_\pi \approx 4 \times 10^{-7} \epsilon_{p,1}$ s, we define the break energy $\epsilon_\mathrm{brk}^{(\pi)}$. Below this energy, pions decay without energy loss, while above this energy the decay spectrum is suppressed by a factor $t_\mathrm{\pi,rad}/\tau_\pi \propto \epsilon_\pi^{-2}$. To take this spectral steepening into account, we define the suppression factor
\begin{eqnarray}
\zeta(\epsilon_\nu) = 
\left\{
\begin{array}{ll}
1 & \quad \mathrm{for}\,\,\epsilon_\nu < \epsilon_{\nu,\mathrm{brk}}^{(\pi)}
\\
( \epsilon_{\nu,\mathrm{brk}}^{(\pi)} / \epsilon_\nu )^{2}
&\quad\mathrm{for}\,\, \epsilon_\nu \ge \epsilon_{\nu,\mathrm{brk}}^{(\pi)},
\end{array} \right.
\end{eqnarray}
where $\epsilon_{\nu,\mathrm{brk}}^{(\pi)} \approx 300$ GeV for our chosen parameters and He star ($\approx$$40$ GeV for the CO star and $\approx$$2 \times 10^4$ GeV for the BSG).

In high-energy proton-proton interactions, $\pi^+$, $\pi^-$, and $\pi^0$ are produced in nearly equal numbers. While inside the star, the density is high enough for the proton-proton optical depth $\tau_{pp} = \int_{R}^{R_*} \mathrm{d}r \, \sigma_{pp} \rho_u(r)/m_p$ to be very high, so that all of the energy in accelerated protons is converted to pions. Thus, we normalize the pion spectrum by the total energy, rather than by the particle number. Since 2/3 of the produced pions are charged, the total energy in charged pions is $E_{\pi^\pm} \approx 2E_p/3$. Here, $E_p$ is the total energy in accelerated protons, which we parameterize as $E_p = \xi_p E_{\rm{exp}}$, where $\xi_p$ is the fraction of the supernova shock energy $E_{\rm{exp}}$ that is channelled into accelerated protons. For strong shocks around supernova remnants, the inferred values of $\xi_p$ are of the order of 0.1 \citep{1987PhR...154....1B,Hillas:2005cs}. In charged pion decay, the muon neutrino takes approximately 1/4 of the pion energy. The flavour ratio of neutrinos produced at the source is $\nu_e^0:\nu_\mu^0:\nu_\tau^0=0:1:0$. Neutrino oscillations en route to a detector on Earth then lead to an observed ratio $\nu_e:\nu_\mu:\nu_\tau \approx 1:1.8:1.8$ \citep{Learned:1994wg,Beacom:2003nh,2003RvMP...75..345G}. The differential fluence of muon neutrinos ($\nu_\mu+\bar{\nu}_\mu$) from a supernova at a distance $D$ is therefore given as
\begin{equation} \label{fluence}
\frac{dF_{\nu}}{d\epsilon_{\nu}} \approx \frac{1}{4 \pi D^2}
\frac{\xi_{\rm{th}} \xi_\nu \xi_p E_{\mathrm{exp}}}
{\mathrm{ln}(\epsilon_{p,\mathrm{max}}/\epsilon_{p,\rm{th}})\epsilon_{\nu}^2}
\zeta(\epsilon_\nu),
\end{equation}
where $\zeta(\epsilon_\nu)$ is the suppression factor due to pion cooling, $\xi_\nu=1/15$, and $\xi_{\rm{th}}$ is the fraction of $E_p$ lying above the pion production threshold energy $\epsilon_{p,\rm{th}}$,
\begin{equation}
\xi_{\rm{th}}=\frac{\int^{\epsilon_{p, \mathrm{max}}}_{\epsilon_{p,
\rm{th}}}d\epsilon_p \epsilon_p dN/dp}
{\int^{\epsilon_{p,\mathrm{max}}}_{m_p}d\epsilon_p \epsilon_p dN/dp}.
\end{equation}

Muon decay also contributes to the neutrino flux, through $\mu^+\to e^+ \bar{\nu}_\mu \nu_e$ and $\mu^- \to e^- \nu_\mu \bar{\nu}_e$. However, due to their smaller mass and longer decay time, they only significantly contribute at lower energies compared to pions. The break energy, determined in the same way as for pions, is $\epsilon_{\nu,\mathrm{brk}}^{(\mu)} \approx 20$ GeV for our chosen parameters and He star ($\approx$$3$ GeV for the CO star and $\approx$$1300$ GeV for the BSG). Note that muon collisional cooling, evaluated using the cross section of \cite{1975ICRC....6.1949B}, is not significant, and muons cool most rapidly by inverse-Compton on electron synchrotron photons. Below the break energy, the total flavour ratio of neutrinos at the source, i.e., combined with muon neutrinos from pion decay, is $\nu_e^0:\nu_\mu^0:\nu_\tau^0=1:2:0$. The observed ratio after oscillations is then $\nu_e:\nu_\mu:\nu_\tau \approx 1:1:1$ \citep{Learned:1994wg}, and $\xi_\nu =1/6$. The change in flavour ratio can also be a probe of this transition \citep{2005PhRvL..95r1101K}. For our purposes, the inclusion of muon decay increases the muon neutrino flux by a factor 5/2, for energies below $\epsilon_{\nu,\mathrm{brk}}^{(\mu)}$.

%------------------------------------------------------------------------------
\subsection{Neutrino-induced muon detection}

In this section we discuss the detection prospects of the neutrinos discussed in previous sections. We address the question of the maximum accelerated proton energy, which determines the maximum neutrino energy and hence detectability at Super-Kamiokande and IceCube detectors. In particular, detection by IceCube is great improved by the IceCube deepcore \citep{Resconi:2008fe}, with its low detection threshold energy of $\epsilon_\nu \sim$10 GeV. 

As the supernova shock propagates outwards, the maximum proton energy increases. In order to address acceleration in the outer envelope, we make use of the analytic formula according to \cite{1999ApJ...510..379M} for the mass density at the edge of the star \citep[see also e.g.,~][]{2003ApJ...584..390W}, 
\begin{equation} \label{EdgeDensity}
\rho^\prime=\rho_* \left( \frac{R_*}{r}-1 \right)^n,
\end{equation}
where $n=(\gamma-1)^{-1}=3$ for a radiative envelope, and $\gamma$ is the adiabatic index. We fit this function to our chosen stellar models. The CO star is approximately described by the parameters $\rho_*=2 \,\mathrm{g \,cm^{-3}}$ and $R_*=5 \times 10^{10} \,\mathrm{cm}$. The He star is described by the parameters $\rho_*=0.01 \,\mathrm{g \,cm^{-3}}$ and $R_*= 2 \times 10^{11}$ cm, while the BSG is described by the parameters $\rho_*= 6 \times 10^{-5} \,\mathrm{g \,cm^{-3}}$ and $R_*=3.4 \times10^{12}$ cm. 

Using the fits to the stellar density, we determine the maximum proton energy at $R_\tau(\tau_{pp})$, the radius defined by a proton-proton optical depth to the stellar surface of $\tau_{pp}$. We use Eq.~(\ref{maxenergy}), which is justified since proton-proton collision is the dominant proton energy loss process. This yields $\epsilon_{p,\mathrm{max}} \sim 8 \times 10^3$ GeV for the fiducial He star at $R_\tau(\tau_{pp}=5)$ ($\sim$$1\times 10^4$ GeV for the CO star and $\sim$$900$ GeV for the BSG). Although larger radii yield larger $\epsilon_{p,\mathrm{max}}$, we do not consider this for several reasons. First, the fraction of accelerated protons interacting with target protons rapidly decrease since the target density falls as a steep function of radius. Although protons can also interact with photons, the required photon energy is very high, $\epsilon_p \epsilon_\gamma > 0.3 \,\mathrm{GeV^2}$. Second, we have normalized the neutrino spectrum under the condition that all accelerated protons lose energy by multiple proton-proton collisions. Third, higher energy neutrino emission is typically strongly suppressed due to pion cooling. Therefore, while larger radii yield higher energy neutrinos, they do not lead to more detected neutrinos.

The total number of $\nu_\mu$-induced muon events in a neutrino detector is the integral over energy of $\rho_t V_\mathrm{det} \sigma_{CC} dF_{\nu}/d\epsilon_{\nu}$, where $\rho_t$ is the target density, $V_\mathrm{det}$ is the detector volume, and $\sigma_{CC}$ is the inelastic neutrino-nucleon cross section. We take into account the muon range \citep[e.g.,][]{Beacom:2003nh}, which effectively increases the detector volume by detecting muons produced outside the instrumented volume. We adopt the cross section used in \cite{Ashie:2005ik}, which covers the energy range down to $\epsilon_\nu \approx 0.1$ GeV. Between $1 < \epsilon_{\nu,\mathrm{GeV}} <10^3$, the cross section increases roughly proportional to $\epsilon_{\nu}$ \citep{Gandhi:1998ri}. The energy of the muon faithfully represents the neutrino energy, and we use the average of the $\nu$-$N$ and $\bar{\nu}_\mu$-$N$ values of $\left< y \right>$ as computed in \cite{Gandhi:1995tf}, where $y(\epsilon_\nu)=1-\epsilon_\mu / \epsilon_\nu$. The opacity of the Earth becomes comparable to 1 when the neutrino energies are $\sim$10 TeV or more; the Earth is therefore totally transparent for the neutrinos of our interest.

We adopt $E_{\rm{exp}}=10^{51}$ erg and distance $D=10$ kpc, i.e., a supernova occurring near the galactic centre. First we discuss detection with Super-Kamiokande. Integrating the fluence over neutrino energy 0.1--$0.05 \epsilon_{p,\mathrm{max}} $ GeV, the expected number of neutrino induced muons is
\begin{equation} \label{signal}
N_\mu \approx 160 \xi_{p,-1}E_{\rm{exp},51} D_{22.5}^{-2},
\end{equation} 
for our fiducial He star. Prospects for detection of the CO star and BSG are similar, with total events of 70 and 130 respectively. Note that the contribution from muon decay is non-negligible at Super-Kamiokande energies, since the muon break energy is above the detection threshold. For example, of the 160 events, we find that $\sim$50 are from muon decays.

For IceCube deepcore, we conservatively estimate the effective area as
\begin{eqnarray}
A_{\mathrm{eff}}(\epsilon_\nu) = 
\left\{
\begin{array}{ll}
2.0 \times 10^{-3} \,(\epsilon_\nu/10)^{3.5} \, \mathrm{cm^2} 
& 10 < \epsilon_\nu \le 10^2
\\
6.3 \, (\epsilon_\nu/10^2)^{1.7} \, \mathrm{cm^2} 
& 10^2 < \epsilon_\nu \le 10^4,
\end{array} \right.
\end{eqnarray} 
which contains neutrino interaction probability, muon propagation, detector response, and event selection \citep[e.g.,][]{Desiati:2006qc,Rott:2008aa}. The neutrino energy is in GeV. We integrate over neutrino energy 10--$0.05 \epsilon_{p,\mathrm{max}} $ GeV, yielding the expected number of neutrino induced muons
\begin{equation}
N_\mu \approx 6600 \xi_{p,-1}E_{\rm{exp},51} D_{22.5}^{-2},
\end{equation} 
for the fiducial He star. Total events for the CO star and BSG are $\sim$600 and $\sim$200 respectively. The suppression in the CO star case is due to the low $\epsilon_{\nu,\mathrm{brk}}^{(\pi)}$, while for the BSG it is the low $\epsilon_{p,\mathrm{max}}$. Note that we have neglected contributions from muon decays for IceCube predictions, which are expected to be small. We also note that the break energy is a function of $\xi_e$, the fraction of total energy in electrons, by $\epsilon_{\nu,\mathrm{brk}}^{(\pi)} \propto \xi_e^{-1/2}$. We have conservatively assumed $\xi_e=0.01$. For example, for $\xi_e=10^{-3}$ \citep{2001ApJ...558..739A,2003ApJ...589..827B}, the break energy for the Type Ic case is $\approx$10$^4$ GeV, and the total event number increases from $\sim$200 to $\sim$3000. 

All the events discussed will cluster in a time window $\approx R_*/v_s$ which is of the order of seconds to hours, depending on the progenitor size. In comparison, the background muon rate due to atmospheric neutrinos, over the entire $2\pi$ steradian sky in 1 day, is $\sim$10 for Super-Kamiokande and $\sim$100 for IceCube, using the same assumptions of detection efficiency as our signal calculations. These are comparable to values in the literature, $\sim$10 for Super-Kamiokande \citep{Kajita:2000mr} and $O$(100) for IceCube \citep{GonzalezGarcia:2005xw}. We add that the true number of signal events will differ by more than a factor of a few, depending on the quality of cuts used. However, any increase or decrease from our estimates will similarly increase or decrease atmospheric background events, and our conclusion that the signal overwhelms background is unaffected. Combined with the angular correlation with an shock breakout \citep{Soderberg:2008uh} or the optical supernova, an essentially background free detection of the neutrino signal is possible, as long as $N_\mu>1$.

%------------------------------------------------------------------------------
\subsection{Diffuse nonthermal neutrino background}

From Eq.~(\ref{fluence}), the emitted number (per unit energy range) of nonthermal neutrinos from each supernova is $dN_{\nu} / d\epsilon_{\nu} = 4 \pi D^2 dF_{\nu} / d\epsilon_{\nu} \approx 1 \times 10^{51} \zeta(\epsilon_{\nu,0}) \epsilon_{\nu,0}^{-2} \, \mathrm{GeV^{-1}}$, adopting fiducial parameter values for the He star model. These neutrinos have been continuously injected to the Universe since the beginning of star formation. Since they are unabsorbed, they form a diffuse nonthermal neutrino background in the present Universe. Thus, it is of interest to estimate the intensity of this diffuse background, and compare it with other diffuse components such as atmospheric neutrinos and diffuse thermal neutrino background from supernovae \citep[e.g.,][]{2003APh....18..307A,2004NJPh....6..170A}.
 
A key element for such an estimate is the occurrence rate of supernovae leaving behind strongly magnetized compact objects in the Galaxy. The relatively young age of the observed magnetars indicate a lower Galactic magnetar birth rate of 1 in $10^4$ years \citep{1994Natur.368..125K,1995A&A...299L..41V}. However, due to low detection efficiencies caused by e.g., on/off states of magnetars, it has been suggested that the rate may be an order larger, approaching the rate of radio pulsars \citep{2006csxs.book..547W}. Therefore, we adopt the magnetar birth rate of $10^{-3}$\,yr$^{-1}$ per galaxy. Given the average galaxy number density, $n_{\rm gal} \simeq 10^{-2}\,\mathrm{Mpc}^{-3}$ \citep{Blanton:2000ns,2003AJ....125.1682N}, the global occurrence rate of supernovae accompanying magnetar remnants is $R_{\rm SN,mag} = 10^{-5} \, \mathrm{Mpc^{-3}} \, \mathrm{yr^{-1}}$. 

The intensity of the diffuse nonthermal neutrino background is therefore roughly estimated as
\begin{eqnarray}
\frac{d\Phi_{\nu}}{d\epsilon_{\nu}} & \approx &
\frac{c}{4\pi}\frac{dN_{\nu}}{d\epsilon_{\nu}} R_{\rm SN,mag}
t_H \xi_{\rm SF}
\nonumber\\ 
&=& 1 \times 10^{-8} \zeta(\epsilon_{\nu,0}) 
\xi_{\rm SF} \epsilon_{\nu,0}^{-2} \,
\mathrm{GeV^{-1} cm^{-2} \, s^{-1} \, \, sr^{-1}},
\end{eqnarray}
where $t_H = 10^{10}$\ yr is the Hubble time, and we neglected the decrease of neutrino energy due to cosmic expansion (i.e., energy redshift) for simplicity. We have optimistically assumed that each magnetar producing supernovae yields neutrino emissions comparable to our He star model. It is natural to assume that the magnetar birth rate, associated with the deaths of short-lived massive stars, is proportional to the cosmic star formation rate; the correction factor $\xi_{\rm SF}$ takes this into account. Since the star formation rate was larger by an order of magnitude in the past Universe at redshifts $\sim$1--2 \citep[e.g.,][]{Hopkins:2006bw}, $\xi_{\rm SF}$ might be about a few \citep[see, for a similar discussion,][]{Waxman:1998yy}.

We compare this intensity with that of atmospheric neutrinos. Around 0.1 GeV, the latter is $\sim 1 \, \mathrm{GeV^{-1} \, cm^{-2} \, s^{-1} \, sr^{-1} }$ \citep{Gaisser:1988ar,Daum:1994bf,Malek:2002ns}. It is best to use the lowest-energy bin, around 0.1 GeV, to maximize the signal-to-noise ratio. At this energy, the intensity is $\sim$$3\times 10^{-6} \xi_{{\rm SF},0.5} \, \mathrm{GeV^{-1} \, cm^{-2} \, s^{-1} \, sr^{-1}}$. This is many orders of magnitude smaller than the intensity of atmospheric neutrinos, which makes it difficult to use the diffuse nonthermal neutrino background for extracting information on model parameters. We also note that this intensity at 0.1 GeV is smaller than the exponential tail of the intensity of diffuse thermal neutrino background from supernovae \citep{2003APh....18..307A,2004NJPh....6..170A}.

%==============================================================================
\section{Discussion and conclusions}

In this paper we investigated proton acceleration and high-energy neutrino emission from the core-collapse supernovae of strongly magnetized stars. The stellar magnetic field assumed in this work is determined under the fossil field hypothesis of magnetar magnetic fields. Since the CO core is generally opaque to high-energy neutrinos \citep[e.g.,~][]{Horiuchi:2007xi}, we consider only high-energy neutrinos from the epoch when the supernova shock propagates through the stellar envelope. We considered three progenitor models and showed that protons may be accelerated up to $\sim$10$^4$ GeV (CO star), $\sim$$10^4$ GeV (He star), and $\sim$$10^3$ GeV (BSG) respectively. These are above the pion production threshold, and we can expect high-energy neutrinos from pion decay. In all progenitors studied, the maximum proton energy is limited by proton-proton collisional cooling. 

The neutrino signal will be detectable above background atmospheric neutrinos by the Super-Kamiokande detector, if a small fraction $> 6\times 10^{-4}$ of the explosion energy is channelled into accelerated protons. Here we have quoted the results for a He star explosion (Type Ib supernova) of $E_\mathrm{exp}=10^{51}$ erg occurring at 10 kpc; this gives $N_\mu>1$ (see Eq.~(\ref{signal})). The required fractions are slightly larger for explosions of CO stars and BSGs. These neutrinos, detected after thermal neutrinos but prior to the optical supernova, act as probes of large stellar magnetic fields. 

The neutrino signal can also be detected by km$^3$ detectors such as IceCube, which are sensitive to higher energy ($\gtrsim 10$ GeV) neutrinos. The predicted total event numbers for IceCube deepcore are $\sim$600 (CO star explosion), $\sim$6600 (He star explosion), and $\sim$200 (BSG explosion) respectively. The strong suppression in a CO star explosion is due to strong pion cooling by inverse-Compton on electron synchrotron photons. Thus the CO star result depends on the total energy of accelerated electrons, which is not well known. We have assumed a large fraction, noting that a smaller value would increase the total event number significantly. On the other hand, the strong suppression in a BSG star explosion is due to a low maximum proton energy.

While the magnetar birth rate is conservatively estimated as 10$^{-4}$\,yr$^{-1}$ in the Galaxy, it could be larger by an order of magnitude because of its detection inefficiency. In addition, of the $\sim$10 magnetar candidates, a few lie within 3 kpc away \citep{Kothes:2002gt,2004A&A...416.1037H,Gaensler:2005qk}, predicting a prolific neutrino signal. Both of these facts are positive aspects for testing this scenario by burst nonthermal neutrino detection. On the other hand, the diffuse nonthermal neutrino background from these sources is much smaller than conventional atmospheric neutrinos.

We have focused on the epoch of supernova shock propagation in the stellar envelope. As the shock continues propagating outwards, it eventually enters optically thin matter, i.e., it crosses the photosphere. At this point, radiation leaks forwards, and the shock makes a transition from radiation mediated to collisional \citep{Ensman:1991td}. The collisional phase may be brief, as investigated by \cite{Waxman:2001kt}. The latter authors find that the growth rate of electromagnetic instabilities within the shocked material is larger than the collisional rate, and the shock becomes collisionless. Emission of TeV neutrinos have been predicted at this stage \citep{Waxman:2001kt}. Our scenario can be distinguished from this emission by lower neutrino energies, as well as the no-detection of simultaneous $\gamma$-rays from neutral pion decay. 

We have for simplicity assumed a spherical supernova shock. Another simplification is our treatment of the magnetic field. In reality, different shock and field geometries (parallel and perpendicular orientations) yield different acceleration efficiencies, and further treatments of both geometries and dynamics are required to address the full spectrum and luminosity curve of the neutrino signal. These are beyond the scope of this paper, but we add that our simple assumption is in part due to the lack of realistic models of the magnetic field inside stars.

Regardless of our approximations, we can show that any detection of the nonthermal neutrinos will occur after thermal neutrinos (if the case of SN 1987A is standard, then $O$(10) hours after), and last $R_*/v_s \approx$ seconds to hours, depending on the progenitor size. The neutrino signature from Type II supernovae, with larger progenitor radii, are expected to be longer in duration compared to those from smaller Type Ibc explosions. Detection will also be feasible for optically dark supernovae in the Galactic Centre or in molecular clouds, which are heavily obscured by dust.

%==============================================================================
\section*{Acknowledgments}

We are grateful to Kohta Murase for useful discussions. This study was supported by a Grants-in-Aid for Scientific Research from the Ministry of Education, Science and Culture of Japan (No. S19104006; SH and KS), Japan Society for Promotion of Science (YS and TH), Sherman Fairchild Foundation (SA), and by World Premier International Research Center InitiativeiWPI Initiative), MEXT, Japan.

%\begin{thebibliography}{99}
 
\bibliography{bibliography}

\begin{thebibliography}{}

\bibitem[\protect\citeauthoryear{{Allen}, {Petre} \& {Gotthelf}}{{Allen}
  et~al.}{2001}]{2001ApJ...558..739A}
{Allen} G.~E.,  {Petre} R.,    {Gotthelf} E.~V.,  2001, \apj, 558, 739

\bibitem[\protect\citeauthoryear{{Ando} \& {Sato}}{{Ando} \&
  {Sato}}{2004}]{2004NJPh....6..170A}
{Ando} S.,  {Sato} K.,  2004, New Journal of Physics, 6, 170

\bibitem[\protect\citeauthoryear{{Ando}, {Sato} \& {Totani}}{{Ando}
  et~al.}{2003}]{2003APh....18..307A}
{Ando} S.,  {Sato} K.,    {Totani} T.,  2003, Astropart. Phys., 18, 307

\bibitem[\protect\citeauthoryear{Ashie et~al.,}{Ashie
  et~al.}{2005}]{Ashie:2005ik}
Ashie Y.,  et~al., 2005, Phys. Rev., D71, 112005

\bibitem[\protect\citeauthoryear{{Baade} \& {Zwicky}}{{Baade} \&
  {Zwicky}}{1934}]{BaadeZwicky1934}
{Baade} W.,  {Zwicky} F.,  1934, Phys. Rev., 46, 76

\bibitem[\protect\citeauthoryear{{Bamba}, {Yamazaki}, {Ueno} \&
  {Koyama}}{{Bamba} et~al.}{2003}]{2003ApJ...589..827B}
{Bamba} A.,  {Yamazaki} R.,  {Ueno} M.,    {Koyama} K.,  2003, \apj, 589, 827

\bibitem[\protect\citeauthoryear{Beacom, Bell, Hooper, Pakvasa \&
  Weiler}{Beacom et~al.}{2003}]{Beacom:2003nh}
Beacom J.~F.,  Bell N.~F.,  Hooper D.,  Pakvasa S.,    Weiler T.~J.,  2003,
  Phys. Rev., D68, 093005

\bibitem[\protect\citeauthoryear{Bellandi, Covolan, Costa, Montanha \&
  Mundim}{Bellandi et~al.}{1994}]{Bellandi:1994dy}
Bellandi J.,  Covolan R. J.~M.,  Costa C. G.~S.,  Montanha J.,    Mundim L.~M.,
   1994, Phys. Rev., D50, 297

\bibitem[\protect\citeauthoryear{{Blandford} \& {Eichler}}{{Blandford} \&
  {Eichler}}{1987}]{1987PhR...154....1B}
{Blandford} R.,  {Eichler} D.,  1987, \physrep, 154, 1

\bibitem[\protect\citeauthoryear{Blanton et~al.,}{Blanton
  et~al.}{2001}]{Blanton:2000ns}
Blanton M.~R.,  et~al., 2001, \apj, 121, 2358

\bibitem[\protect\citeauthoryear{Borog \& Petrukhin}{Borog \&
  Petrukhin}{1975}]{1975ICRC....6.1949B}
Borog V.~V.,  Petrukhin A.~A.,  1975, in Proceedings of the 14th International
  Cosmic Ray Conference, Vol.~6.
pp 1949--1954

\bibitem[\protect\citeauthoryear{Brenner et~al.,}{Brenner
  et~al.}{1982}]{Brenner:1981kf}
Brenner A.~E.,  et~al., 1982, Phys. Rev., D26, 1497

\bibitem[\protect\citeauthoryear{Daum et~al.,}{Daum
  et~al.}{1995}]{Daum:1994bf}
Daum K.,  et~al., 1995, Z. Phys., C66, 417

\bibitem[\protect\citeauthoryear{Desiati}{Desiati}{2006}]{Desiati:2006qc}
Desiati P.,  2006, astro-ph/0611603

\bibitem[\protect\citeauthoryear{{Donati}, {Babel}, {Harries}, {Howarth},
  {Petit} \& {Semel}}{{Donati} et~al.}{2002}]{2002MNRAS.333...55D}
{Donati} J.-F.,  {Babel} J.,  {Harries} T.~J.,  {Howarth} I.~D.,  {Petit} P.,
   {Semel} M.,  2002, \mnras, 333, 55

\bibitem[\protect\citeauthoryear{{Donati}, {Howarth}, {Bouret}, {Petit},
  {Catala} \& {Landstreet}}{{Donati} et~al.}{2006}]{2006MNRAS.365L...6D}
{Donati} J.-F.,  {Howarth} I.~D.,  {Bouret} J.-C.,  {Petit} P.,  {Catala} C.,
   {Landstreet} J.,  2006, \mnras, 365, L6

\bibitem[\protect\citeauthoryear{{Donati}, {Howarth}, {Jardine}, {Petit},
  {Catala}, {Landstreet}, {Bouret}, {Alecian}, {Barnes}, {Forveille}, {Paletou}
  \& {Manset}}{{Donati} et~al.}{2006}]{2006MNRAS.370..629D}
{Donati} J.-F.,  {Howarth} I.~D.,  {Jardine} M.~M.,  {Petit} P.,  {Catala} C.,
  {Landstreet} J.~D.,  {Bouret} J.-C.,  {Alecian} E.,  {Barnes} J.~R.,
  {Forveille} T.,  {Paletou} F.,    {Manset} N.,  2006, \mnras, 370, 629

\bibitem[\protect\citeauthoryear{{Drury}}{{Drury}}{1983}]{1983RPPh...64.1429T}
{Drury} L.~O.~C.,  1983, Reports of Progress in Physics, 46, 973

\bibitem[\protect\citeauthoryear{{Duncan} \& {Thompson}}{{Duncan} \&
  {Thompson}}{1992}]{1992ApJ...392L...9D}
{Duncan} R.~C.,  {Thompson} C.,  1992, \apjl, 392, L9

\bibitem[\protect\citeauthoryear{Eidelman et~al.,}{Eidelman
  et~al.}{2004}]{Eidelman:2004wy}
Eidelman S.,  et~al., 2004, Phys. Lett., B592, 1

\bibitem[\protect\citeauthoryear{Ensman \& Burrows}{Ensman \&
  Burrows}{1992}]{Ensman:1991td}
Ensman L.,  Burrows A.,  1992, \apj, 393, 742

\bibitem[\protect\citeauthoryear{{Ferrario} \& {Wickramasinghe}}{{Ferrario} \&
  {Wickramasinghe}}{2005}]{2005MNRAS.356..615F}
{Ferrario} L.,  {Wickramasinghe} D.,  2005, \mnras, 356, 615

\bibitem[\protect\citeauthoryear{{Ferrario} \& {Wickramasinghe}}{{Ferrario} \&
  {Wickramasinghe}}{2006}]{2006MNRAS.367.1323F}
{Ferrario} L.,  {Wickramasinghe} D.,  2006, \mnras, 367, 1323

\bibitem[\protect\citeauthoryear{Gaensler et~al.,}{Gaensler
  et~al.}{2005}]{Gaensler:2005qk}
Gaensler B.~M.,  et~al., 2005, \apj, 620, L95

\bibitem[\protect\citeauthoryear{{Gaensler}, {Slane}, {Gotthelf} \&
  {Vasisht}}{{Gaensler} et~al.}{2001}]{2001ApJ...559..963G}
{Gaensler} B.~M.,  {Slane} P.~O.,  {Gotthelf} E.~V.,    {Vasisht} G.,  2001,
  \apj, 559, 963

\bibitem[\protect\citeauthoryear{Gaisser, Stanev \& Barr}{Gaisser
  et~al.}{1988}]{Gaisser:1988ar}
Gaisser T.~K.,  Stanev T.,    Barr G.,  1988, Phys. Rev., D38, 85

\bibitem[\protect\citeauthoryear{Gandhi, Quigg, Reno \& Sarcevic}{Gandhi
  et~al.}{1996}]{Gandhi:1995tf}
Gandhi R.,  Quigg C.,  Reno M.~H.,    Sarcevic I.,  1996, Astropart. Phys., 5,
  81

\bibitem[\protect\citeauthoryear{Gandhi, Quigg, Reno \& Sarcevic}{Gandhi
  et~al.}{1998}]{Gandhi:1998ri}
Gandhi R.,  Quigg C.,  Reno M.~H.,    Sarcevic I.,  1998, Phys. Rev., D58,
  093009

\bibitem[\protect\citeauthoryear{{Gavriil}, {Kaspi} \& {Woods}}{{Gavriil}
  et~al.}{2002}]{2002Natur.419..142G}
{Gavriil} F.~P.,  {Kaspi} V.~M.,    {Woods} P.~M.,  2002, \nat, 419, 142

\bibitem[\protect\citeauthoryear{Gonzalez-Garcia, Halzen \&
  Maltoni}{Gonzalez-Garcia et~al.}{2005}]{GonzalezGarcia:2005xw}
Gonzalez-Garcia M.~C.,  Halzen F.,    Maltoni M.,  2005, Phys. Rev., D71,
  093010

\bibitem[\protect\citeauthoryear{{Harding} \& {Lai}}{{Harding} \&
  {Lai}}{2006}]{2006RPPh...69.2631H}
{Harding} A.~K.,  {Lai} D.,  2006, Reports of Progress in Physics, 69, 2631

\bibitem[\protect\citeauthoryear{Hillas}{Hillas}{2005}]{Hillas:2005cs}
Hillas A.~M.,  2005, J. Phys., G31, R95

\bibitem[\protect\citeauthoryear{Hopkins \& Beacom}{Hopkins \&
  Beacom}{2006}]{Hopkins:2006bw}
Hopkins A.~M.,  Beacom J.~F.,  2006, \apj, 651, 142

\bibitem[\protect\citeauthoryear{Horiuchi \& Ando}{Horiuchi \&
  Ando}{2008}]{Horiuchi:2007xi}
Horiuchi S.,  Ando S.,  2008, Phys. Rev., D77, 063007

\bibitem[\protect\citeauthoryear{{Hulleman}, {van Kerkwijk} \&
  {Kulkarni}}{{Hulleman} et~al.}{2004}]{2004A&A...416.1037H}
{Hulleman} F.,  {van Kerkwijk} M.~H.,    {Kulkarni} S.~R.,  2004, \aap, 416,
  1037

\bibitem[\protect\citeauthoryear{{Ibrahim}, {Swank} \& {Parke}}{{Ibrahim}
  et~al.}{2003}]{2003ApJ...584L..17I}
{Ibrahim} A.~I.,  {Swank} J.~H.,    {Parke} W.,  2003, \apjl, 584, L17

\bibitem[\protect\citeauthoryear{Kajita \& Totsuka}{Kajita \&
  Totsuka}{2001}]{Kajita:2000mr}
Kajita T.,  Totsuka Y.,  2001, Rev. Mod. Phys., 73, 85

\bibitem[\protect\citeauthoryear{{Kashti} \& {Waxman}}{{Kashti} \&
  {Waxman}}{2005}]{2005PhRvL..95r1101K}
{Kashti} T.,  {Waxman} E.,  2005, Physical Review Letters, 95, 181101

\bibitem[\protect\citeauthoryear{{Kitaura}, {Janka} \& {Hillebrandt}}{{Kitaura}
  et~al.}{2006}]{2006A&A...450..345K}
{Kitaura} F.~S.,  {Janka} H.-T.,    {Hillebrandt} W.,  2006, \aap, 450, 345

\bibitem[\protect\citeauthoryear{Kothes, Uyaniker \& Yar}{Kothes
  et~al.}{2002}]{Kothes:2002gt}
Kothes R.,  Uyaniker B.,    Yar A.,  2002, \apj, 576, 169

\bibitem[\protect\citeauthoryear{{Kouveliotou}, {Dieters}, {Strohmayer}, {van
  Paradijs}, {Fishman}, {Meegan}, {Hurley}, {Kommers}, {Smith}, {Frail} \&
  {Murakami}}{{Kouveliotou} et~al.}{1998}]{1998Natur.393..235K}
{Kouveliotou} C.,  {Dieters} S.,  {Strohmayer} T.,  {van Paradijs} J.,
  {Fishman} G.~J.,  {Meegan} C.~A.,  {Hurley} K.,  {Kommers} J.,  {Smith} I.,
  {Frail} D.,    {Murakami} T.,  1998, \nat, 393, 235

\bibitem[\protect\citeauthoryear{{Kouveliotou}, {Fishman}, {Meegan},
  {Paciesas}, {van Paradijs}, {Norris}, {Preece}, {Briggs}, {Horack},
  {Pendleton} \& {Green}}{{Kouveliotou} et~al.}{1994}]{1994Natur.368..125K}
{Kouveliotou} C.,  {Fishman} G.~J.,  {Meegan} C.~A.,  {Paciesas} W.~S.,  {van
  Paradijs} J.,  {Norris} J.~P.,  {Preece} R.~D.,  {Briggs} M.~S.,  {Horack}
  J.~M.,  {Pendleton} G.~H.,    {Green} D.~A.,  1994, \nat, 368, 125

\bibitem[\protect\citeauthoryear{Learned \& Pakvasa}{Learned \&
  Pakvasa}{1995}]{Learned:1994wg}
Learned J.~G.,  Pakvasa S.,  1995, Astropart. Phys., 3, 267

\bibitem[\protect\citeauthoryear{{Longair}}{{Longair}}{1994}]{1981Longair}
{Longair} M.~S.,  1994, High Energy Astrophysics, second edition edn.
Cambridge University Press, Cambridge

\bibitem[\protect\citeauthoryear{Malek et~al.,}{Malek
  et~al.}{2003}]{Malek:2002ns}
Malek M.,  et~al., 2003, Phys. Rev. Lett., 90, 061101

\bibitem[\protect\citeauthoryear{{Malkov} \& {Drury}}{{Malkov} \&
  {Drury}}{2001}]{2001RPPh...64.1429T}
{Malkov} M.~A.,  {Drury} L.~O.~C.,  2001, Reports of Progress in Physics, 64,
  429

\bibitem[\protect\citeauthoryear{{Matzner} \& {McKee}}{{Matzner} \&
  {McKee}}{1999}]{1999ApJ...510..379M}
{Matzner} C.~D.,  {McKee} C.~F.,  1999, \apj, 510, 379

\bibitem[\protect\citeauthoryear{{Mazzali}, {Deng}, {Nomoto}, {Sauer}, {Pian},
  {Tominaga}, {Tanaka}, {Maeda} \& {Filippenko}}{{Mazzali}
  et~al.}{2006}]{2006Natur.442.1018M}
{Mazzali} P.~A.,  {Deng} J.,  {Nomoto} K.,  {Sauer} D.~N.,  {Pian} E.,
  {Tominaga} N.,  {Tanaka} M.,  {Maeda} K.,    {Filippenko} A.~V.,  2006, \nat,
  442, 1018

\bibitem[\protect\citeauthoryear{{Moss}}{{Moss}}{2003}]{2003A&A...403..693M}
{Moss} D.,  2003, \aap, 403, 693

\bibitem[\protect\citeauthoryear{{Nakamura}, {Fukugita}, {Yasuda}, {Loveday},
  {Brinkmann}, {Schneider}, {Shimasaku} \& {SubbaRao}}{{Nakamura}
  et~al.}{2003}]{2003AJ....125.1682N}
{Nakamura} O.,  {Fukugita} M.,  {Yasuda} N.,  {Loveday} J.,  {Brinkmann} J.,
  {Schneider} D.~P.,  {Shimasaku} K.,    {SubbaRao} M.,  2003, AJ, 125, 1682

\bibitem[\protect\citeauthoryear{{Paczynski}}{{Paczynski}}{1992}]{1992AcA....4%
2..145P}
{Paczynski} B.,  1992, Acta Astronomica, 42, 145

\bibitem[\protect\citeauthoryear{{Petit}, {Wade}, {Drissen}, {Montmerle} \&
  {Alecian}}{{Petit} et~al.}{2008}]{2008MNRAS.387L..23P}
{Petit} V.,  {Wade} G.~A.,  {Drissen} L.,  {Montmerle} T.,    {Alecian} E.,
  2008, \mnras, 387, L23

\bibitem[\protect\citeauthoryear{{Rea}, {Israel}, {Stella}, {Oosterbroek},
  {Mereghetti}, {Angelini}, {Campana} \& {Covino}}{{Rea}
  et~al.}{2003}]{2003ApJ...586L..65R}
{Rea} N.,  {Israel} G.~L.,  {Stella} L.,  {Oosterbroek} T.,  {Mereghetti} S.,
  {Angelini} L.,  {Campana} S.,    {Covino} S.,  2003, \apjl, 586, L65

\bibitem[\protect\citeauthoryear{Resconi}{Resconi}{2008}]{Resconi:2008fe}
Resconi E.,  {for the IceCube Collaboration,} 2008, in Status and prospects of
  the IceCube neutrino telescope, arXiv:0807.3891

\bibitem[\protect\citeauthoryear{Rott}{Rott}{2008}]{Rott:2008aa}
Rott C.,  {for the IceCube Collaboration,} 2008, in Proceedings of the 34th
  International Conference on High Energy Physics, arXiv:0810.3698

\bibitem[\protect\citeauthoryear{{Sim{\'o}n-D{\'{\i}}az}, {Herrero}, {Esteban}
  \& {Najarro}}{{Sim{\'o}n-D{\'{\i}}az} et~al.}{2006}]{2006A&A...448..351S}
{Sim{\'o}n-D{\'{\i}}az} S.,  {Herrero} A.,  {Esteban} C.,    {Najarro} F.,
  2006, \aap, 448, 351

\bibitem[\protect\citeauthoryear{Soderberg et~al.,}{Soderberg
  et~al.}{2008}]{Soderberg:2008uh}
Soderberg A.~M.,  et~al., 2008, Nature., 453, 469

\bibitem[\protect\citeauthoryear{{Thompson} \& {Duncan}}{{Thompson} \&
  {Duncan}}{1993}]{1993ApJ...408..194T}
{Thompson} C.,  {Duncan} R.~C.,  1993, \apj, 408, 194

\bibitem[\protect\citeauthoryear{{Thompson} \& {Duncan}}{{Thompson} \&
  {Duncan}}{1995}]{1995MNRAS.275..255T}
{Thompson} C.,  {Duncan} R.~C.,  1995, \mnras, 275, 255

\bibitem[\protect\citeauthoryear{{Thompson} \& {Duncan}}{{Thompson} \&
  {Duncan}}{1996}]{1996ApJ...473..322T}
{Thompson} C.,  {Duncan} R.~C.,  1996, \apj, 473, 322

\bibitem[\protect\citeauthoryear{{Thompson} \& {Duncan}}{{Thompson} \&
  {Duncan}}{2001}]{2001ApJ...561..980T}
{Thompson} C.,  {Duncan} R.~C.,  2001, \apj, 561, 980

\bibitem[\protect\citeauthoryear{{Toma}, {Ioka}, {Sakamoto} \&
  {Nakamura}}{{Toma} et~al.}{2007}]{2007ApJ...659.1420T}
{Toma} K.,  {Ioka} K.,  {Sakamoto} T.,    {Nakamura} T.,  2007, \apj, 659, 1420

\bibitem[\protect\citeauthoryear{{Tout}, {Wickramasinghe} \& {Ferrario}}{{Tout}
  et~al.}{2004}]{2004MNRAS.355L..13T}
{Tout} C.~A.,  {Wickramasinghe} D.,    {Ferrario} L.,  2004, \mnras, 355, L13

\bibitem[\protect\citeauthoryear{{Usov}}{{Usov}}{1992}]{1992Natur.357..472U}
{Usov} V.~V.,  1992, \nat, 357, 472

\bibitem[\protect\citeauthoryear{{van Paradijs}, {Taam} \& {van den
  Heuvel}}{{van Paradijs} et~al.}{1995}]{1995A&A...299L..41V}
{van Paradijs} J.,  {Taam} R.~E.,    {van den Heuvel} E.~P.~J.,  1995, \aap,
  299, L41

\bibitem[\protect\citeauthoryear{Waxman \& Bahcall}{Waxman \&
  Bahcall}{1999}]{Waxman:1998yy}
Waxman E.,  Bahcall J.~N.,  1999, Phys. Rev., D59, 023002

\bibitem[\protect\citeauthoryear{Waxman \& Loeb}{Waxman \&
  Loeb}{2001}]{Waxman:2001kt}
Waxman E.,  Loeb A.,  2001, Phys. Rev. Lett., 87, 071101

\bibitem[\protect\citeauthoryear{{Waxman} \& {M{\'e}sz{\'a}ros}}{{Waxman} \&
  {M{\'e}sz{\'a}ros}}{2003}]{2003ApJ...584..390W}
{Waxman} E.,  {M{\'e}sz{\'a}ros} P.,  2003, \apj, 584, 390

\bibitem[\protect\citeauthoryear{{Weaver}}{{Weaver}}{1976}]{1976ApJS...32..233%
W}
{Weaver} T.~A.,  1976, \apjs, 32, 233

\bibitem[\protect\citeauthoryear{{Wickramasinghe} \&
  {Ferrario}}{{Wickramasinghe} \& {Ferrario}}{2005}]{2005MNRAS.356.1576W}
{Wickramasinghe} D.,  {Ferrario} L.,  2005, \mnras, 356, 1576

\bibitem[\protect\citeauthoryear{{Woods} \& {Thompson}}{{Woods} \&
  {Thompson}}{2006}]{2006csxs.book..547W}
{Woods} P.~M.,  {Thompson} C.,  2006, {Soft gamma repeaters and anomalous X-ray
  pulsars: magnetar candidates}.
Compact stellar X-ray sources, pp 547--586

\bibitem[\protect\citeauthoryear{{Woosley} \& {Heger}}{{Woosley} \&
  {Heger}}{2006}]{2006ApJ...637..914W}
{Woosley} S.~E.,  {Heger} A.,  2006, \apj, 637, 914

\end{thebibliography}
\bibliographystyle{mn2e} 

%\end{thebibliography}

\appendix

\bsp

\label{lastpage}

\end{document}